\documentclass[11pt]{article}
\usepackage{moriond,epsfig}

\newcommand{\be}{\begin{equation}}
\newcommand{\ee}{\end{equation}}
\newcommand{\bi}{\begin{itemize}}
\newcommand{\ei}{\end{itemize}}
\newcommand{\bc}{\begin{center}}
\newcommand{\ec}{\end{center}}
\newcommand{\bea}{\begin{eqnarray}}
\newcommand{\eea}{\end{eqnarray}}

\def    \raw           {\rightarrow}

\def    \part          {\partial}

\begin{document}
\begin{flushright}
{IFT-UAM/CSIC-03-51} \\
{\tt hep-ph/0407009}
\end{flushright}
\vspace{1cm}

\title{Why care about $(\theta_{13},\delta)$ degeneracy at future neutrino experiments}

\author{ S. Rigolin }

\address{Departamento de Fisica Teorica y IFT, Universidad Autonoma de
 Madrid, Madrid, Spain}

\maketitle\abstracts{
The presence of several clone solutions in the simultaneous measurement of 
($\theta_{13},\delta$) has been widely discussed in literature. Here, after 
a pedagogical introduction on why these clones arise, we discuss how the 
clones location in the ($\theta_{13},\delta$) plane change as a function of the 
physical input pair ($\bar\theta_{13},\bar\delta$). We compare the clone flow 
of a set of possible future neutrino experiments: the CERN SuperBeam, BetaBeam 
and Neutrino Factory proposals. We show that the combination of these specific 
BetaBeam and SuperBeam could not help in solving the degeneracies. The combination 
of one of them with the Neutrino Factory Golden and Silver channel can, instead, 
be used to solve completely the eightfold degeneracy.}

\section{Introduction}
\label{sec:intro}

The atmospheric and solar sector of the PMNS leptonic mixing matrix have been measured 
with quite good resolution by SK, SNO and KamLand. These experiments measure two angles, 
$\theta_{12}$ and $\theta_{23}$, and two mass differences, $\Delta m^2_{12}$ and 
$\Delta m^2_{23}$ (for the explicit form of the PMNS matrix and the adopted conventions, 
see for example \cite{Donini:1999jc}). The present bound on $\theta_{13}$, $\sin^2 
\theta_{13} \leq 0.02$, is extracted from the negative results of CHOOZ and from three-family 
analysis of atmospheric and solar data. The PMNS phase $\delta$ is totally unbounded as 
no experiment is sensitive to the leptonic CP violation. The main goal of next neutrino 
experiments will be to measure these two still unknown parameters. In this talk we 
concentrate on analyzing, from a theoretical point of view, the problem of 
degenerations that arise when trying to measure simultaneously ($\theta_{13},\delta$).

\section{The Intrinsic Clone}
\label{sec:intrinsic}

\begin{figure}[t]
\bc
\begin{tabular}{cc}
\hspace{-0.5cm}
\epsfig{file=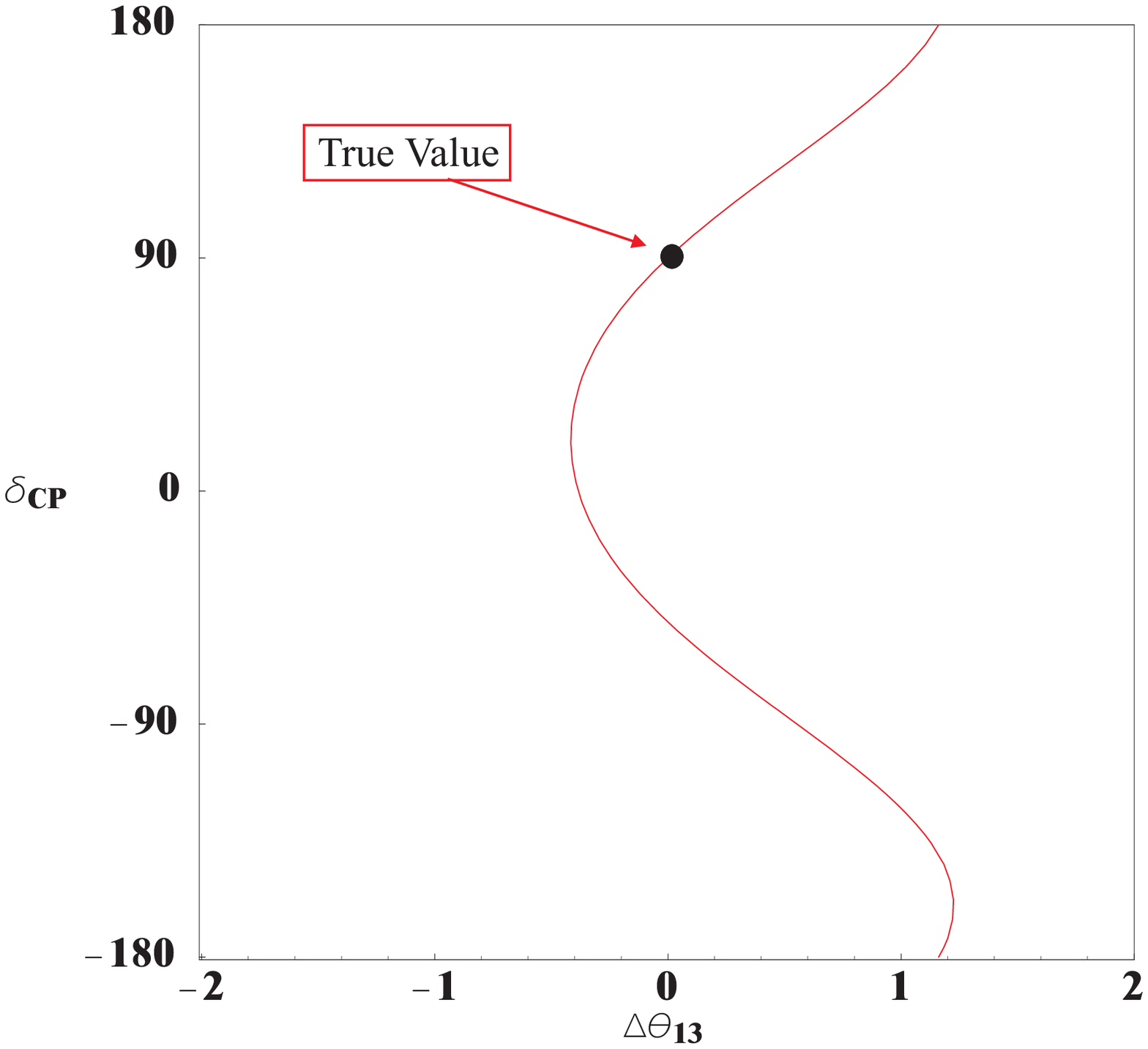,width=7.5cm,angle=0} &
\hspace{-0.5cm}
\epsfig{file=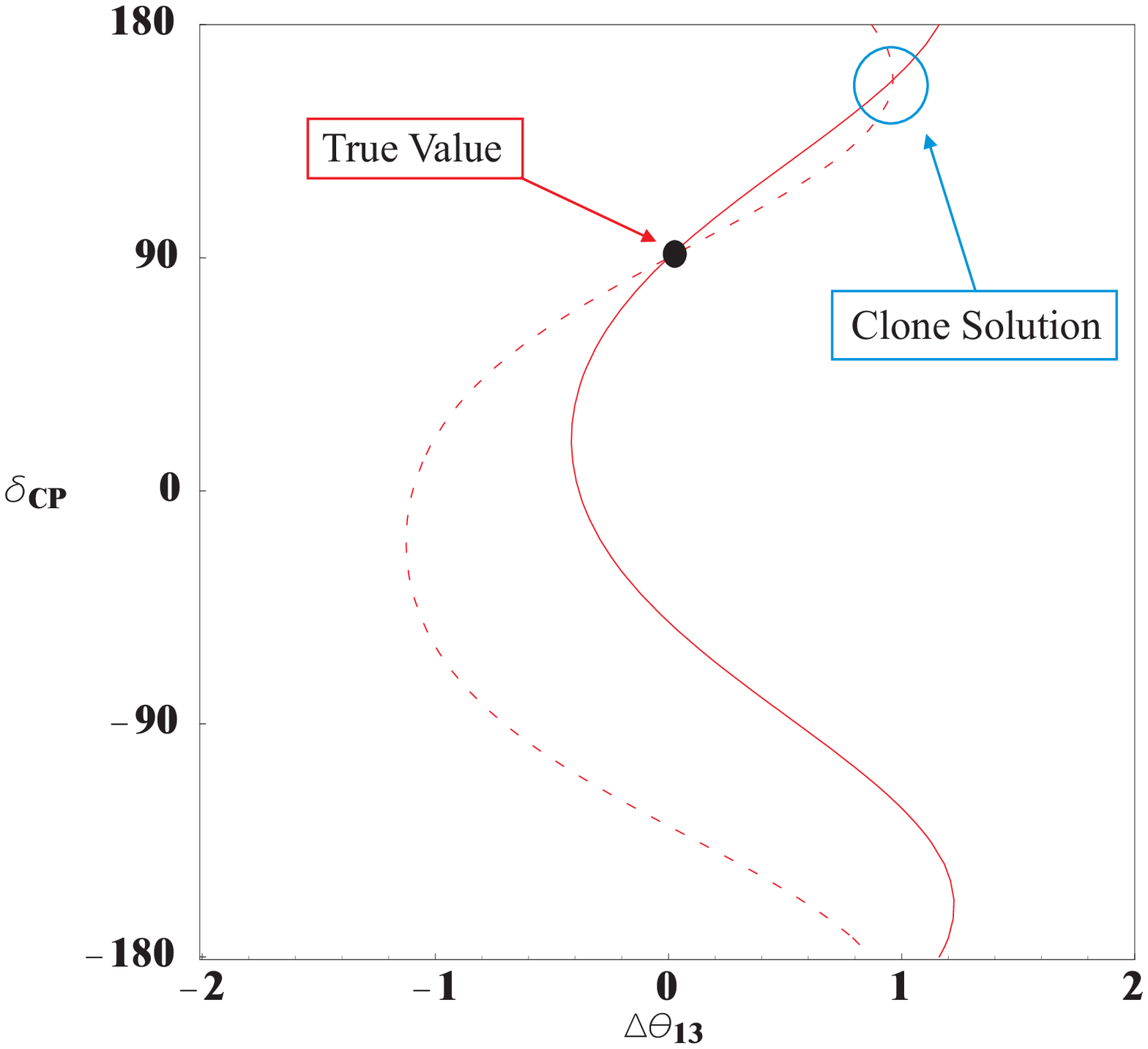,width=7.5cm,angle=0} 
\end{tabular}
\ec
\caption{Clone solutions obtained solving the systems of Eq.~\ref{eq:equi0} 
(left plot) and Eq.~\ref{eq:equi1} (right plot).} 
\label{fig:int1}
\end{figure}

In \cite{Burguet-Castell:2001ez} it has been noticed that the appearance probability 
$P_{\alpha \beta} (\bar \theta_{13},\bar \delta)$ obtained for neutrinos at a fixed 
energy and baseline with input parameter ($\bar\theta_{13}, \bar\delta$) has no unique 
solution. Indeed, the equation:
\be
\label{eq:equi0} 
P_{\alpha \beta} (\bar \theta_{13}, \bar \delta)  =  P_{\alpha \beta} 
(\theta_{13}, \delta)
\ee 
has a continuous number of solutions. The locus of ($\theta_{13},\delta$) satisfying this 
equation is called ``equiprobability curve'', as can be seen in Fig.~\ref{fig:int1} (left). 
Considering the equiprobability curves for neutrinos and antineutrinos with the same energy 
(and the same input parameters), the system of equations ($\pm$ referring to neutrinos and 
antineutrinos respectively):
\be
\label{eq:equi1} 
P^\pm_{\alpha \beta} (\bar \theta_{13},\bar \delta) = P^\pm_{\alpha \beta} 
(\theta_{13}, \delta)
\ee 
has two intersections: the input pair ($\bar \theta_{13},\bar \delta$) and a second, energy 
dependent, point. As can be noticed in Fig.~\ref{fig:int1} (right) this second intersection 
introduces an ambiguity in the measurement of the physical values of $\theta_{13}$ and 
$\delta$: the so-called {\it intrinsic clone} solution.

\begin{figure}[t]
\bc
\begin{tabular}{cc}
\hspace{-0.5cm}
\epsfig{file=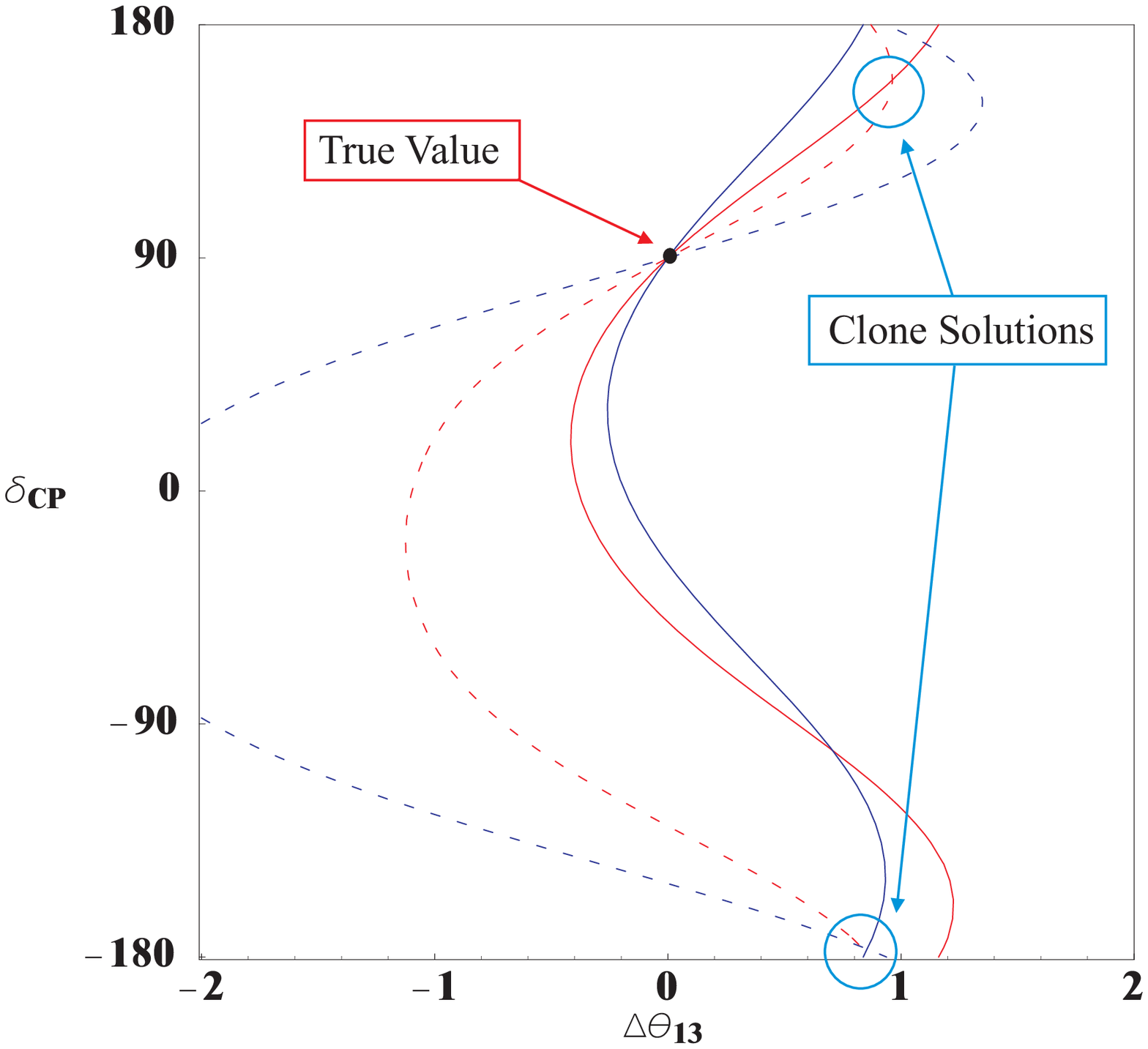,width=7.5cm,angle=0} &
\hspace{-0.5cm}
\epsfig{file=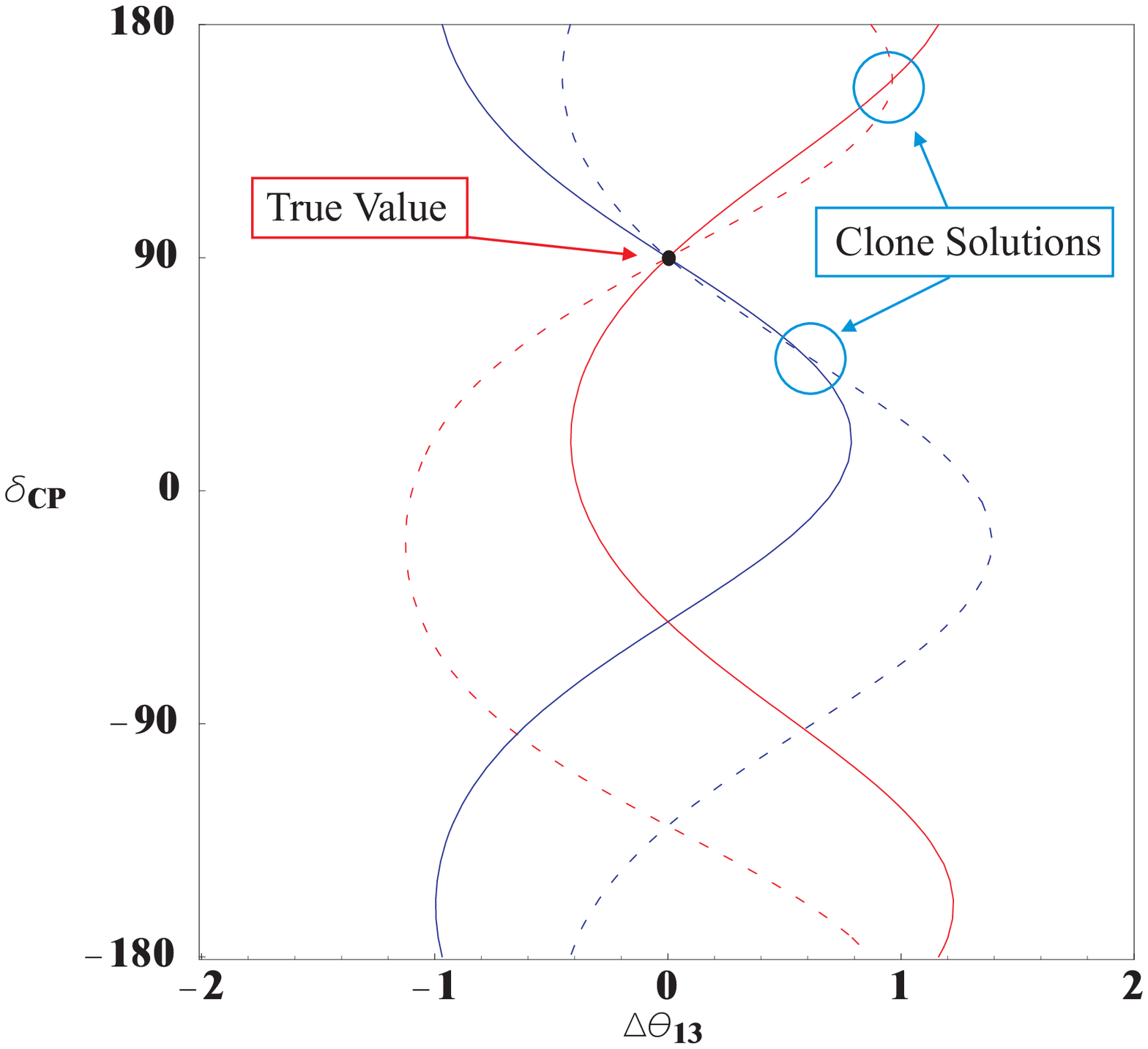,width=7.5cm,angle=0} 
\end{tabular}
\ec
\caption{Solving the intrinsic degeneracy: two baseline L=730 and 3500 km, same channel 
example on the left, vs two channels $\nu_e \raw \nu_\mu$ and $\nu_e \raw \nu_\tau$, same 
baseline example on the right.} 
\label{fig:int2}
\end{figure}

Knowing the two probabilities of Eq.~\ref{eq:equi1} is consequently not enough for 
solving the intrinsic degeneracy. One needs to add more information. Two ways are 
viable: i) using different ``experiments'' (same oscillation channel) and/or ii) 
using different oscillation ``channels'' (at the same experiment). In case i) one can 
think to observe the same neutrino oscillation channel (i.e. the golden $\nu_e \raw 
\nu_\mu$ oscillation) at detectors located at different baseline L. In the left-side 
plot of Fig. \ref{fig:int2} one can see that experiments at different L present clone 
solutions in different regions of the ($\theta_{13},\delta$) parameter space. If the 
clones are well separated one can solve the degeneracy. The same result can be obtained 
if, for the specific neutrino flux considered, bins with quite different energy are 
available\footnote{Like for example in the case of the neutrino factory. This is an 
obvious consequence of the L/E dependence of the degeneracy position.}. Another 
possibility, case ii), is to send the neutrino flux towards only one facility, but to 
use contemporaneously two different oscillation channels (like for example $\nu_e\raw 
\nu_\mu$ and $\nu_e \raw \nu_\tau$). In the right-side plot of Fig. \ref{fig:int2} one 
can see how also in this case the intrinsic clones are expected in different locations 
and the degeneracy can be solved. 

From this example we learn that the best way for solving the degeneracies is to add all 
the possible available information, i.e different baselines, different energy bins and 
different channels. Therefore, in planning future experiments one has to understand which 
combinations of experiments can give the largest set of (really) independent information. 
The existence of unresolved degeneracies could, in fact, manifests itself in a complete 
lost of predictability on the aforementioned parameters.

\section{The Eightfold Degeneracy}
\label{sec:eightfold}

Unfortunately, the appearance of the intrinsic degeneracy is only a part of the ``clone 
problem''. As it was made clear in \cite{Minakata:2001qm,Barger:2001yr} two other sources 
of ambiguities arise due to the present (and probably near future) ignorance of the sign 
of the atmospheric mass difference, $s_{atm}=sign [\Delta m^2_{23}]$ and the $\theta_{23}$ 
octant, $s_{oct}=sign [\tan (2\theta_{23})]$. These two discrete variables assume 
the values $\pm 1$, depending on the physical assignments of the $\Delta m^2_{23}$ sign 
($s_{atm}=1$ for $m_3^2>m_2^2$ and $s_{atm}=-1$ for $m_3^2<m_2^2$) and of the $\theta_{23}$ 
octant ($s_{oct}=1$ for $\theta_{23}<\pi/4$ and $s_{oct}=-1$ for $\theta_{23}>\pi/4$).
As a consequence, future experiments will have as ultimate goal the measure of the two 
continuous variables $\theta_{13}$ and $\delta$ plus the two discrete variables $s_{atm}$ 
and $s_{oct}$. 

Before following with the analysis of the clones position, it should be noticed that 
experimental results are not given in terms of oscillation probabilities but of number of 
charged leptons observed in a specific detector. We must therefore integrate the 
oscillation probability over the neutrino flux, the $\nu N$ cross-section and the 
detector efficiency $\epsilon (E_\mu)$. Eq.~(\ref{eq:equi1}) is thus replaced by
\be
\label{eq:ene0int} 
N^\pm_{\beta} (\bar \theta_{13},\bar \delta; \bar s_{atm},\bar s_{oct}) =
N^\pm_{\beta} (\theta_{13},\delta; s_{atm}=\bar s_{atm}; s_{oct}=\bar s_{oct})\, ,
\ee
with $\beta$ the lepton flavour corresponding to the oscillated neutrino.
In Eq.~(\ref{eq:ene0int}) we have implicitly assumed to know the right sign and the right 
octant for the atmospheric mass difference and angle. As these quantities are unknown 
(and they will remain so in the near future) the following systems of equations should 
be considered as well: 
\bea
\label{eq:ene0sign}
N^\pm_{\beta}(\bar \theta_{13}, \bar \delta; \bar s_{atm}, \bar s_{oct} )&=&
N^\pm_{\beta} ( \theta_{13}, \delta; s_{atm} = -\bar s_{atm}, s_{oct} = \bar
s_{oct})\\  
\label{eq:ene0t23}
N^\pm_{\beta}(\bar \theta_{13}, \bar \delta; \bar s_{atm}, \bar s_{oct}) &=& 
N^\pm_{\beta} ( \theta_{13},  \delta; s_{atm} = \bar s_{atm}, s_{oct} = -\bar
s_{oct})\\ 
\label{eq:ene0t23sign}
N^\pm_{\beta}(\bar \theta_{13}, \bar \delta; \bar s_{atm}, \bar s_{oct} )&=& 
N^\pm_{\beta} ( \theta_{13},  \delta; s_{atm} = -\bar s_{atm}, s_{oct} = -\bar s_{oct})
\eea
Solving the four systems of Eqs.~(\ref{eq:ene0int})-(\ref{eq:ene0t23sign}) will result 
in obtaining the true solution plus the appearance of additional {\it clones} to form an 
eightfold-degeneracy \cite{Barger:2001yr}. These eight solutions are: 
\begin{itemize}
\item the true solution and its {\em intrinsic clone}, obtained solving the system 
      of Eq.~(\ref{eq:ene0int});
\item the $\Delta m^2_{23}$-sign clones (hereafter called {\em sign clones}) of the 
      true and intrinsic solution, obtained solving the system of Eq.~(\ref{eq:ene0sign});
\item the $\theta_{23}$-octant clones (hereafter called {\em octant clones}) of the 
      true and intrinsic solution, obtained solving the system of Eq.~(\ref{eq:ene0t23});
\item the $\Delta m^2_{atm}$-sign $\theta_{23}$-octant clones (hereafter called 
      {\em mixed clones}) of the true and intrinsic solution, obtained solving the system 
      of Eq.~(\ref{eq:ene0t23sign}).
\end{itemize}
It has been noticed\cite{Donini:2003vz} that clones location calculated starting from the 
probability or the number of events can be significantly different.

\section{The Degeneracies flows}
\label{sec:flow}

The analytical description of the clones flows is beyond the non-expert cut of this talk 
and can be found in \cite{Donini:2003vz}. Here we will apply those results to illustrate 
with a specific example, how studying the clones location can give us a hint on which 
combination of experiments is better suited to solve some (or all) of the degeneracies. 
It must be noticed that this analysis only provides a useful tool to detect the experiment 
synergies that are most promising to measure these two still unknown entries of the PMNS 
matrix. These results must then be confirmed by a detailed analysis in which statistics 
and systematics of a given experiment combination are carefully taken into account. 

The experiments that we have considered in our analysis are the following: 
\begin{itemize}
\item A Neutrino Factory with 50 GeV muons circulating in the storage ring and two detectors: 
      the first one located at $L=2810$ Km is designed to look for the golden channel 
      (hereafter labeled as NFG); the second one located at $L=732$ Km is designed to 
      look for the silver channel (hereafter labeled as NFS). The average neutrino energy 
      is: $<E_{\nu_e,\bar \nu_e}> \simeq 30$ GeV.
\item A SuperBeam with 2 GeV protons and one detector located at $L=130$ Km to look for 
      $\nu_\mu \to \nu_e$ (hereafter labeled as SB). The average neutrino energy is: 
      $<E_{\nu_\mu}> \simeq 270$ MeV and $<E_{\bar \nu_\mu}> \simeq 250$ MeV.
\item A Beta Beam with ${}^6$He and ${}^{18}$Ne ions and one detector located at $L=130$ 
      Km to look for $\nu_e \to \nu_\mu$ (hereafter labeled as BB). The average neutrino 
      energy is: $<E_{\nu_e}> \simeq 370$ MeV and $<E_{\bar \nu_e}> \simeq 230$ MeV. 
\end{itemize}

Also if for our theoretical analysis the details of the detectors (such as the mass or 
the specific technology adopted to look for a given signal) are not fundamentals, the 
relevant parameters (the neutrino fluxes and the baseline) have been chosen considering 
the following proposed experiments:  (NFG) \cite{Cervera:2000kp,Cervera:2000vy}, (NFS) 
\cite{Donini:2002rm,Autiero:2003fu}, (SB) \cite{Gomez-Cadenas:2001eu} and (BB) 
\cite{Zucchelli:sa}. For the different beams we have taken as representative the 
CERN Neutrino Factory, SPL and BetaBeam proposals\footnote{See for example in 
\cite{Apollonio:2002en} for a detailed description of each of these proposals and 
\cite{Burguet-Castell:2003vv} for a description of a higher energy BB proposal.}. 

The theoretical requirement for solving ambiguities is to find a specific combination 
of experiments such that the corresponding clone flows lie ``well apart''. In this 
scenario an experimental fit to the data will result in a good $\chi^2$ absolute minimum 
near the true solution, where the information from different experiments adds coherently. 
Near the clone locations poor $\chi^2$ relative minima will be found, since here the 
results of different experiments do not add coherently. 

%

In the following subsections we analyze the flows of the different degeneracies for the 
four considered facilities. We compute the theoretical clone locations for a variable 
$\bar \theta_{13}$ in the range $\bar \theta_{13} \in [0.1^\circ,10^\circ]$, going from 
the present CHOOZ upper bound $[\sin^2 2 \bar \theta_{13}]_{max} = {\cal O} (10^{-1})$ 
down to $[\sin^2 2\bar\theta_{13}]_{min}={\cal O} (10^{-5})$. We show here due to the 
limited space available only the $\bar\delta=90^\circ$ case corresponding to the maximal 
leptonic CP-violation phase. 

\subsection{The Intrinsic and the Sign Clone Flow}
\begin{figure}[t!]
\vspace{-0.5cm}
\begin{center}
\begin{tabular}{cc} 
\hspace{-0.55cm} \epsfxsize7.5cm\epsffile{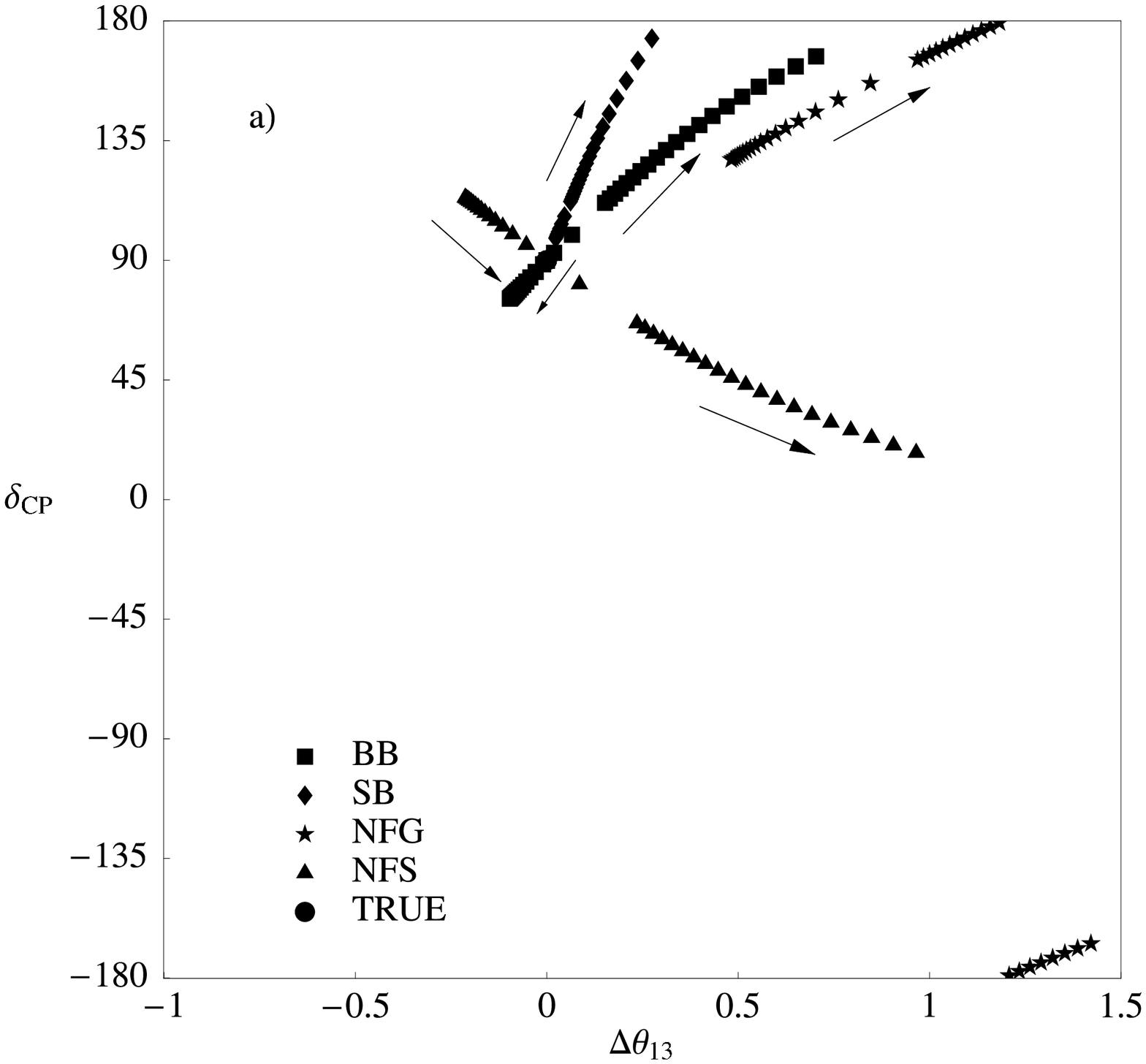} \hspace{-0.3cm} & \hspace{-0.3cm} 
                 \epsfxsize7.5cm\epsffile{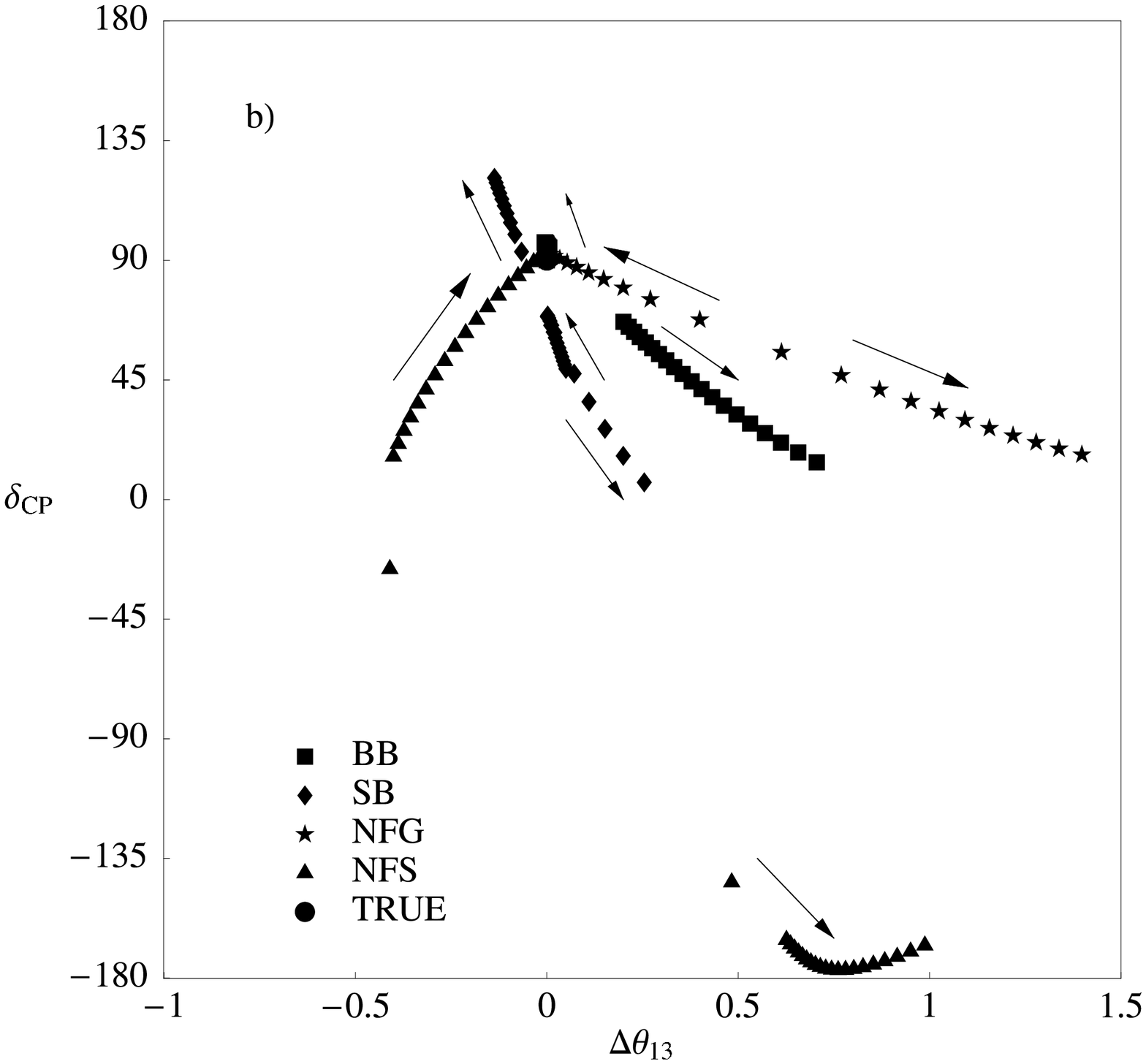} \vspace{-0.5cm}\\ 
\hspace{-0.55cm} \epsfxsize7.5cm\epsffile{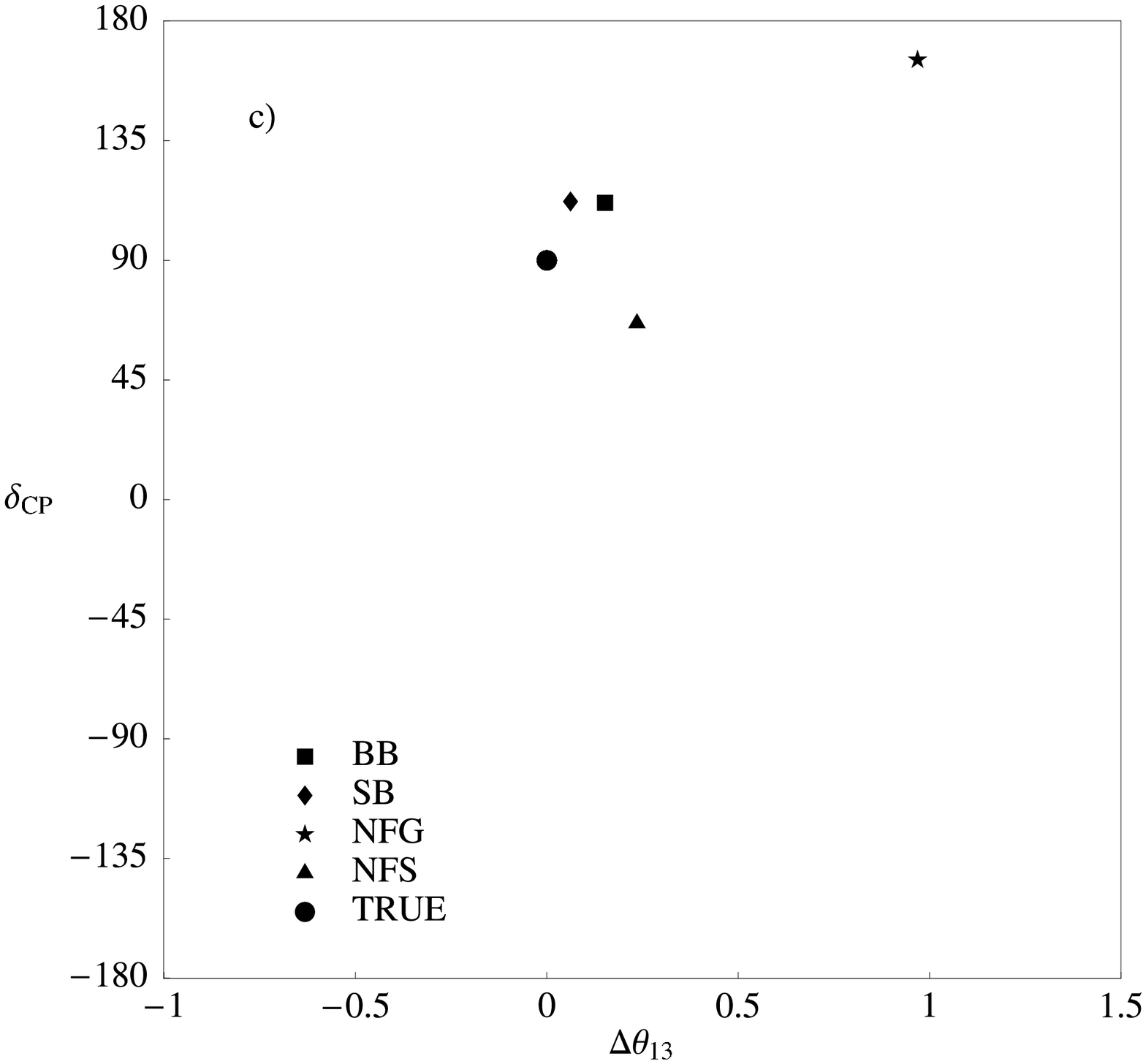} \hspace{-0.3cm} & \hspace{-0.3cm} 
                 \epsfxsize7.5cm\epsffile{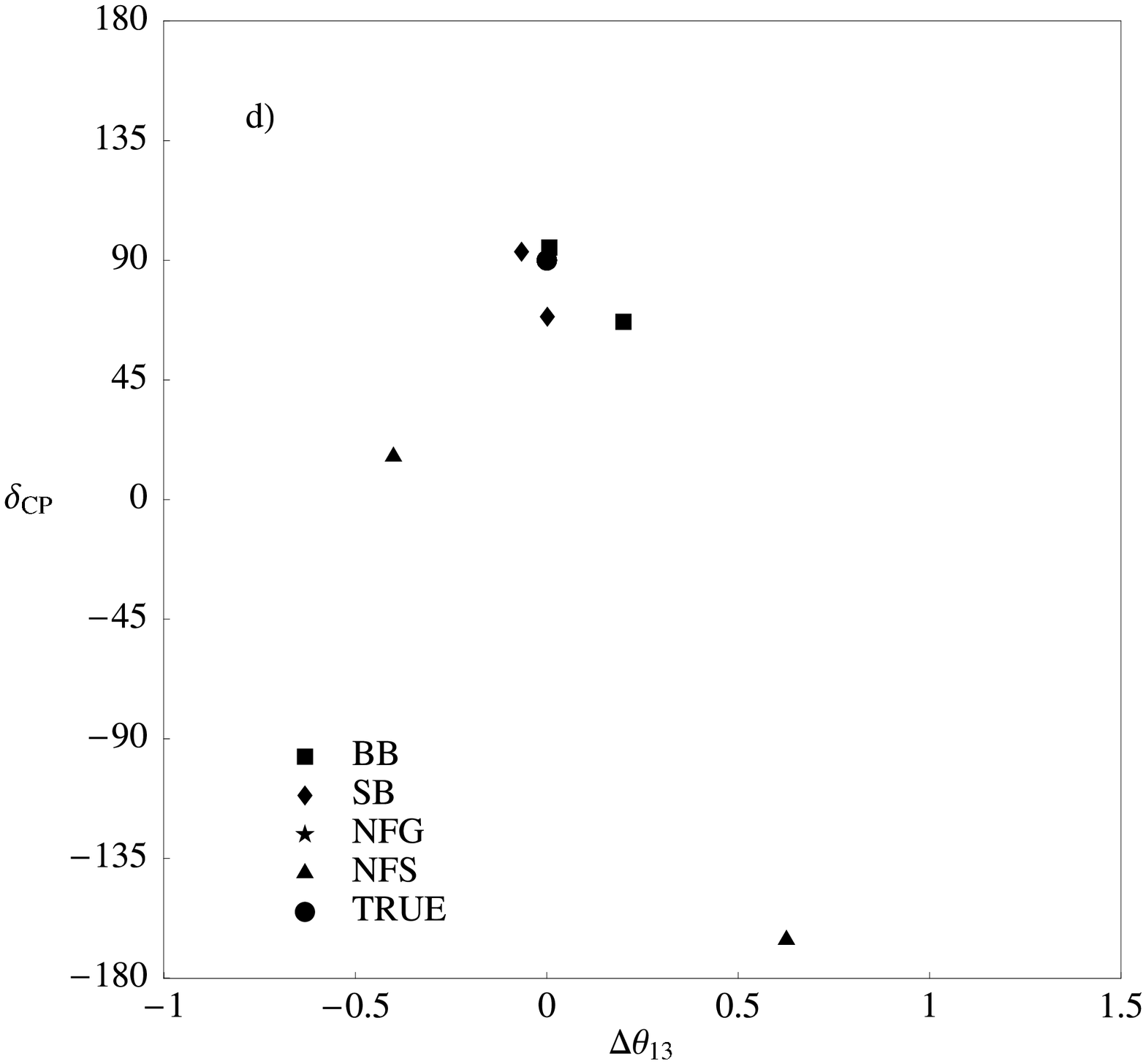} \\
\end{tabular}
\vspace{-0.5cm}
\caption{\it 
Clone location in the ($\Delta\theta_{13},\delta$) plane for the intrinsic (left) and sign 
(right) degeneracies: BB (boxes); SB (diamonds); NFG (stars) and NFS (triangles). The full 
circle is the true solution. (a) and (b) represent the intrinsic and sign clone flows for 
$\bar\theta_{13} \in [0.1^\circ,10^\circ]$ and $\bar\delta=90^\circ$; (c) and (d) represent 
the case $\bar\theta_{13}=1^\circ$.}
\label{fig:allflows1}
\end{center}
\vspace{-0.5cm}
\end{figure}

\vspace{-0.2cm}
In Fig.~\ref{fig:allflows1} we present the clone flow for the intrinsic (left) and the sign 
(right) degeneracies. The clone locations have been computed for $\bar\theta_{13}$ in the 
range $\bar \theta_{13} \in [0.1^\circ,10^\circ]$ and $\bar\delta=90^\circ$ (upper plots) 
or for $\bar \theta_{13} = 1^\circ, \bar \delta = 90^\circ$ (lower plots), on the verge of 
the sensitivity limit. The arrows indicate the direction of the flows from large to small 
$\bar\theta_{13}$. In each figure the results for the four facilities are presented together, 
in order to show how the combination of two or more of them can help solving that specific 
degeneracy. As it can be seen the BB and SB clones both for the intrinsic and the sign 
degeneracy lies very near, making these two experiments not the best choice for resolving 
these degeneracies even if in the specific case shown in the lower plots, the loss of 
predictivity on $\theta_{13}$ and $\delta$ seems ``acceptable''. Conversely NFG or NFS clones 
lie in a well separated region. The golden channel combined with the silver and/or the SB/BB 
facilities is always, in principle, capable to solve the intrinsic ambiguity, while the golden 
channel alone can solve the sign ambiguity (at least for $\theta_{13} \ge 1^\circ$).  

\subsection{The Octant and the Mixed Clones}
\begin{figure}[t!]
\vspace{-0.5cm}
\begin{center}
\begin{tabular}{cc} 
\hspace{-0.55cm} \epsfxsize7.5cm\epsffile{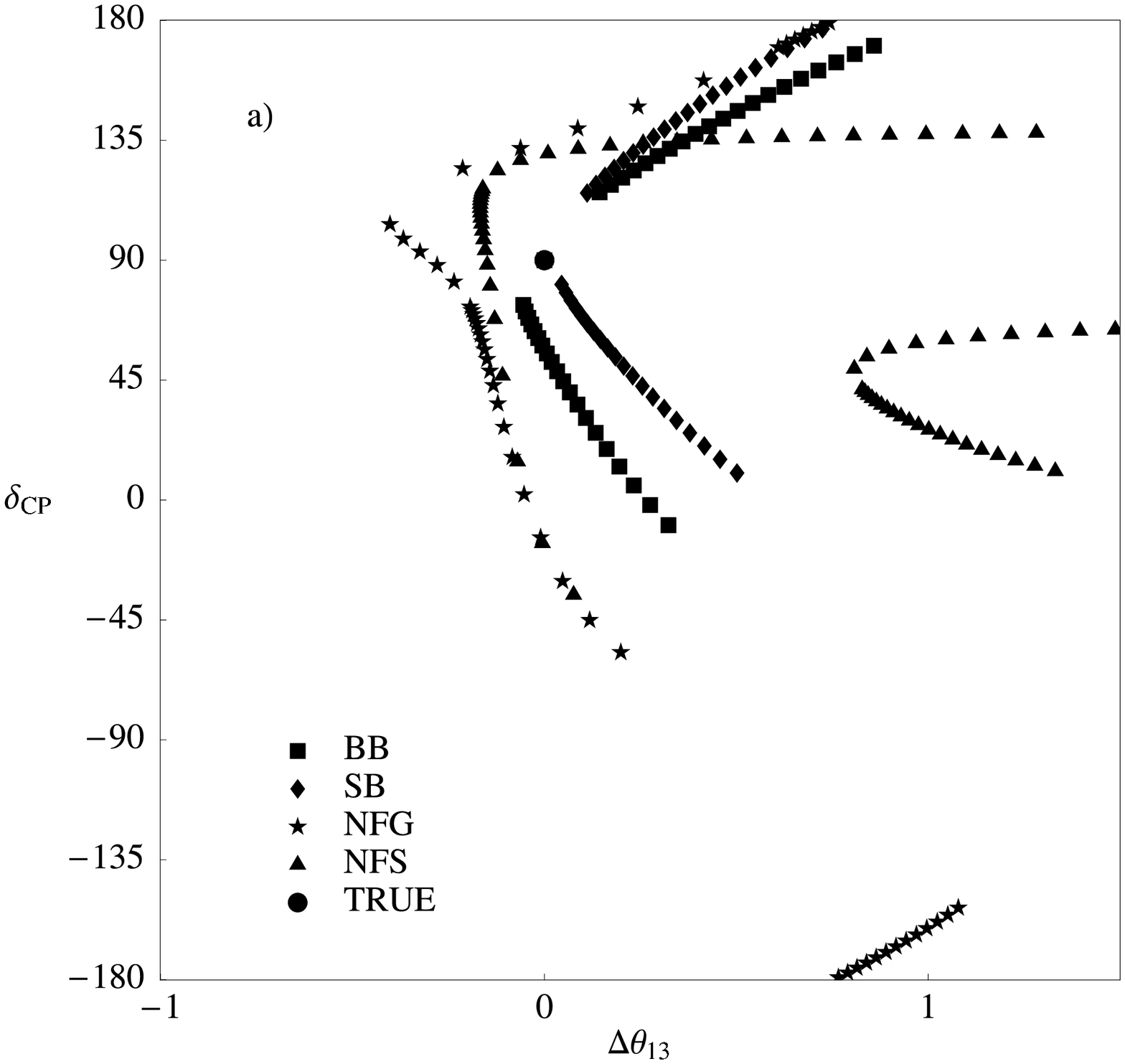} \hspace{-0.3cm} & \hspace{-0.3cm} 
                 \epsfxsize7.5cm\epsffile{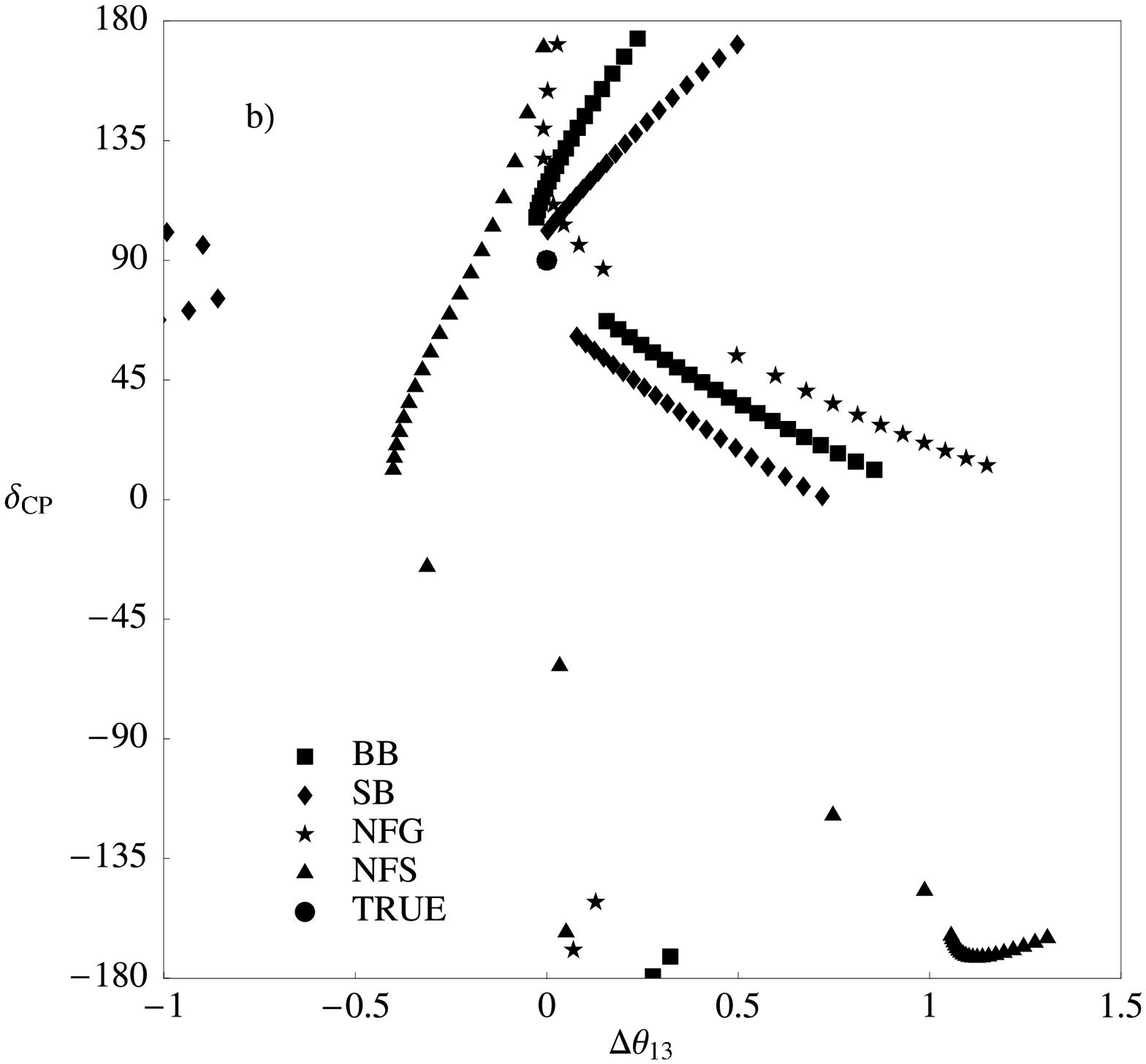} \vspace{-0.5cm} \\
\hspace{-0.55cm} \epsfxsize7.5cm\epsffile{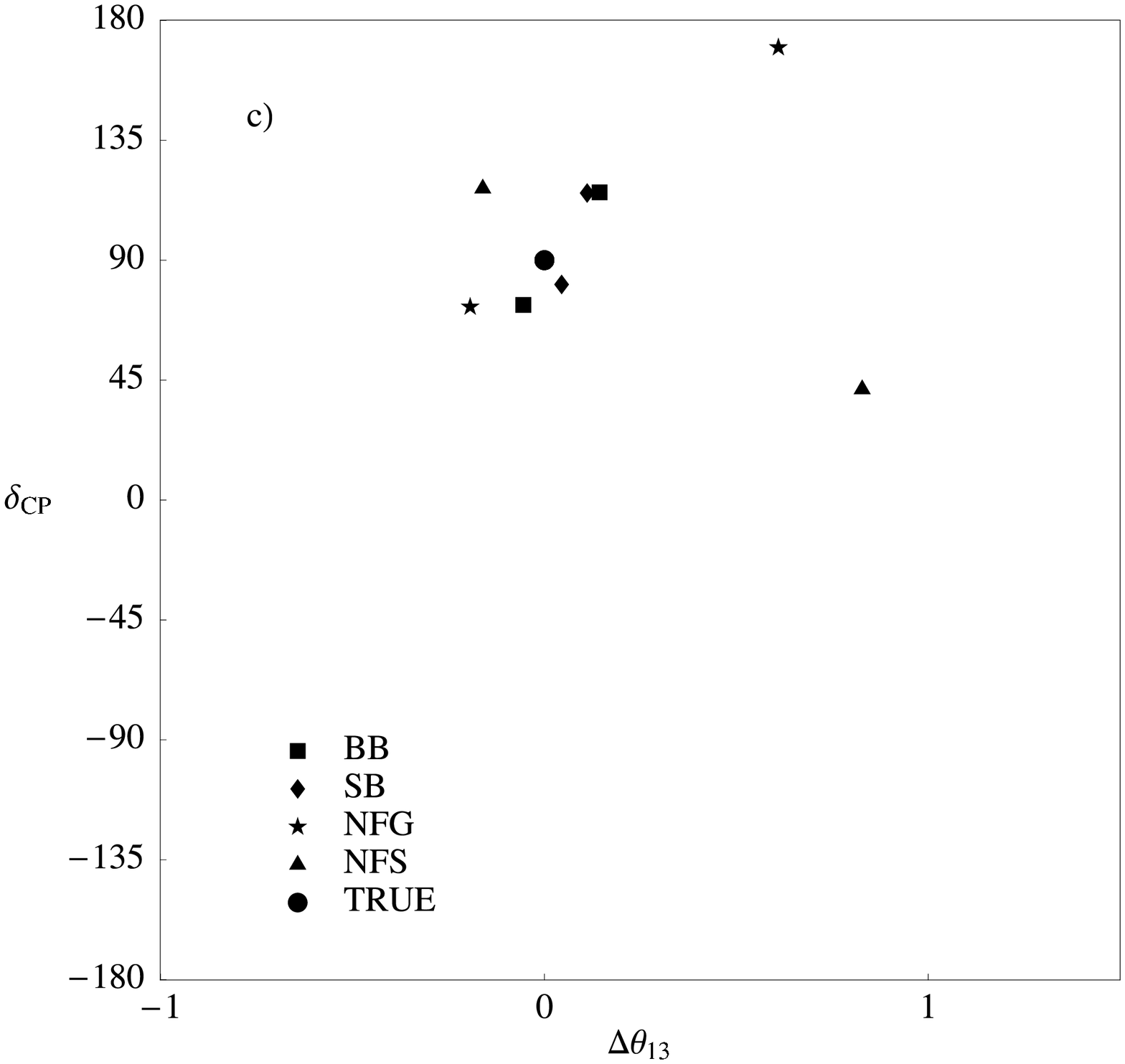} \hspace{-0.3cm} & \hspace{-0.3cm} 
                 \epsfxsize7.5cm\epsffile{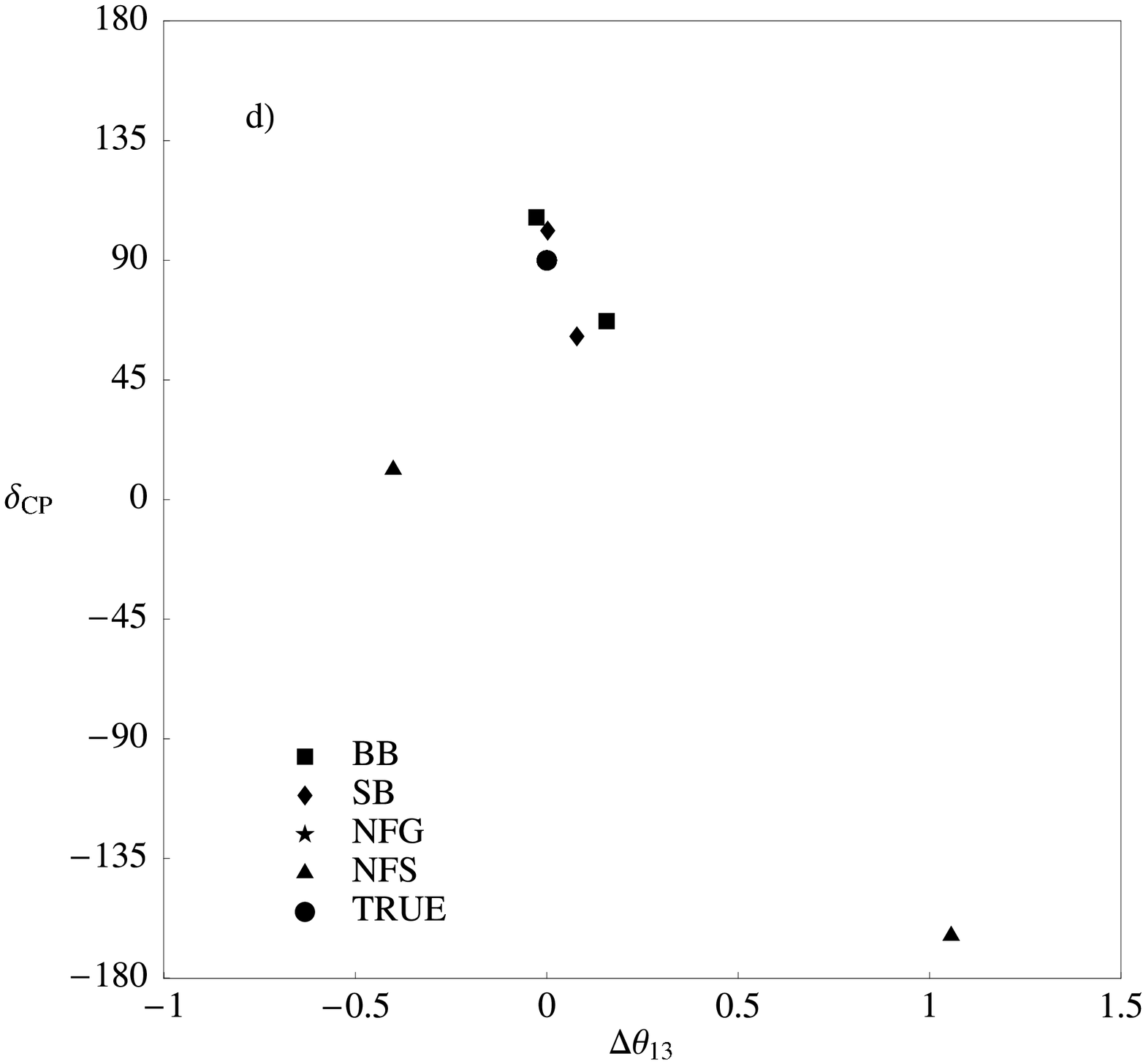} \\
\end{tabular}
\vspace{-0.5cm}
\caption{\it 
Clone location in the ($\Delta\theta_{13},\delta$) plane for the octant (left) and mixed 
(right) degeneracies: BB (boxes); SB (diamonds); NFG (stars) and NFS (triangles). The 
full circle is the true solution. (a) and (b) represent the octant and mixed clone flows 
for $\bar\theta_{13} \in [0.1^\circ,10^\circ]$ and $\bar\delta=90^\circ$; (c) and (d) 
represent the case $\bar\theta_{13}=1^\circ$. The results have been computed with 
$\theta_{23} = 40^\circ$.}
\label{fig:allflows2}
\end{center}
\vspace{-0.5cm}
\end{figure}

\vspace{-0.2cm}
If $\theta_{23}$ were maximal no additional degeneracies would be present. Otherwise two 
further degeneracies do appear. In Fig.~\ref{fig:allflows2} we present the clone flow for 
the octant (left) and the mixed (right) degeneracies. The clone locations have been computed 
for $\bar\theta_{13}$ in the range $\bar \theta_{13} \in [0.1^\circ, 10^\circ]$ and $\bar
\delta=90^\circ$ (upper plot) or for a fixed value, $\bar\theta_{13}=1^\circ, \bar\delta 
= 90^\circ$ (lower plots). In each figure the results for the four facilities are presented 
together.
Again we notice that the BB and SB flows for the octant and mixed degeneracies lie 
always very near making difficult the task of solving degeneracies using only these 
two facilities. This seems a general consequence of the quite similar beam design 
(baseline and flux). NFG and NFS flows are instead always well separated except for 
particularly small $\theta_{13}$ values when some superposition can be observed. 
Anyway using the Neutrino Factory plus on of the SB/BB seems possible, in principle, 
to solve all the degeneracies. NFG alone can solve the mixed degeneracy being sensitive 
to the sign of $\Delta m^2_{23}$ (at least for $\theta_{13} \ge 1^\circ$). 
  
\section{Conclusions}
\label{sec:concl}

In this talk we have analyzed, from a theoretical point of view, the problem of 
degenerations that arise when trying to measure simultaneously ($\theta_{13},\delta$). 
The existence of unresolved degeneracies could, in fact, manifests itself in a complete 
loss of predictability on the aforementioned parameters at future neutrino facilities. 
Therefore, in planning future experiments will be important to understand which 
combinations of experiments can give the largest set of (really) independent information. 
With the setup considered we noticed that BetaBeam/SuperBeam combination seems less 
effective than the Neutrino Factory/BetaBeam (SuperBeam) combination to solve the 
eightfold degeneracy.

\vspace{-0.25cm}
\section*{Acknowledgments}
This work was done in collaboration with A. Donini and D. Meloni. This work was 
supported in part by the Spanish DGI of the MCYT under contract FPA2003-04597.

\end{document}